\documentclass[aps,prb,showpacs,twocolumn,superscriptaddress,floatfix]{revtex4-2}

\usepackage[colorlinks,bookmarks=false,citecolor=blue,linkcolor=red,urlcolor=blue]{hyperref}
\usepackage{epsfig}
\usepackage{hhline}
\usepackage{tikz}
\usepackage{graphicx,comment}
\usepackage{enumitem}
\usepackage{dcolumn}
\usepackage{bm}

\usepackage{amsmath,amssymb,mathtools}
\usepackage{placeins}
\usepackage{bbold}

\usepackage{scalerel}
\makeatletter
\newcommand{\hourglass}{\mathbin{\scalerel*{\@hgpic}{\ensuremath{\sigma}}}}
\newcommand{\@hgpic}{
    \setlength{\unitlength}{0.25cm}
    \begin{picture}(1,1.1)
    \thicklines
    \put(0,0){\line(2,3){1}}
    \put(1,1.5){\line(-1,0){1}}
    \put(0,1.5){\line(2,-3){1}}
    \put(1,0){\line(-1,0){1}}
    \end{picture}
}
\makeatother

\begin{document}
	
	\title{Trivalent network model for d$^3$ transition metal dichalcogenides in the 1T structure: \\Holography from local constraints}
	\author{Ashland Knowles}
	\email{gk23dp@brocku.ca}
	\affiliation{Department of Physics, Brock University, St. Catharines, Ontario L2S 3A1, Canada}
	\author{R. Ganesh}
	\email{r.ganesh@brocku.ca}
	\affiliation{Department of Physics, Brock University, St. Catharines, Ontario L2S 3A1, Canada}
	
	\date{\today}
	
	\begin{abstract}
                Dimer models are well known as prototypes for locally constrained physics. They describe systems in which every site on a lattice must be attached to one dimer. Loop models are an extension of this idea, with the constraint that two dimers must touch at each site. Here, we present a further generalization where every site must have three dimers attached -- a trivalent network model. As concrete physical realizations, we discuss d$^3$ transition metal dichalcogenides in the 1T structure -- materials with the structural formula MX$_2$ (M = Tc, Re) or AM$'$X$_2$ (A = Li or Na; M$'$ = Mo, W), where X is a chalcogen atom. These materials have a triangular layer of transition metal atoms, each with three valence electrons in $t_{2g}$ orbitals. Each atom forms valence bonds with three of its nearest neighbours. The geometry of the 1T structure imbues each bond with sharp orbital character. We argue that this enforces a ``bending constraint'' so that two dimers attached to the same site cannot be parallel. This leads to a highly structured space of configurations, with alternating bonds along each line of the underlying triangular lattice. There is no dynamics, as constraints forbid local  rearrangements of dimers. We construct a phase diagram, identifying configurations that minimize potential energy. We find a rhombus-stripe phase that explains a distortion pattern seen across several materials. Remarkably, the local constraints in this model lead to a simple example of holography. The bonding configuration in the bulk is completely determined by the configuration at the boundary. We recast the model in terms of three Ising chains that are defined on the boundaries of a triangular cluster. As a testable prediction, we propose that a single impurity will generate long-ranged domain walls. 
	\end{abstract}
	
	\maketitle  
	
	\section{Introduction}
A paradigmatic approach for physical systems with local constraints is the dimer model, where dimers are placed on bonds of a lattice such that each site is connected to one dimer. Classical dimer models\cite{Lieb1967,Huse2003,Alet2005,Wilkins2021} may take into account repulsion (or attraction) between proximate dimers. Quantum models\cite{MoessnerRaman2011} include dimer rearrangements that are consistent with the local constraints. Dimer models have helped demonstrate exactly solvable limits\cite{Rokhsar1988}, $\mathbb{Z}_2$ spin liquids\cite{Moessner2001}, deconfined criticality\cite{Vishwanath2004,Alet2006}, etc. A natural extension of this idea is to loop models where two dimers must touch at each site. Loop models have been explored in the context of honeycomb lattice materials\cite{Savary2021}, Rydberg atoms\cite{Ran2024} and transition metal dichalcogenides\cite{Knowles2025}. In this article, we introduce a further extension of this general idea: trivalent network models, where each site is constrained to touch \textit{three} dimers.

We place our discussion within the context of transition metal dichalcogenides (TMDs)\cite{Manzeli2017,Yin2021,Dai2024}. The building block of these materials is an X-M-X trilayer, where M and X represent triangular-lattice layers of transition metal and chalcogen atoms respectively. In the 1T structure, these layers are arranged in an ABC configuration. Each transition metal atom sits within an octahedral cage formed by the chalcogens. Materials with this structure show several distortions. In a previous study\cite{Knowles2025}, the present authors have proposed a quantum loop model for the d$^2$ subfamily of materials, where each transition metal atom has two valence electrons. In this article, we propose a network model for the d$^3$ subfamily. We seek to explain structural distortions, attributing them to patterned arrangements of valence bonds.

An intriguing feature of our model is that it can be recast as an effective lower-dimensional theory. Bond arrangements in the two-dimensional bulk are entirely determined by configurations of three one-dimensional Ising chains. These chains can be viewed as bond variables at the extremities of a cluster, leading to a holographic bulk-boundary mapping.
The effective theory exhibits long-ranged couplings across the Ising chains, reflecting the fact that couplings originate from throughout the bulk of the sample.
The notion of bulk-boundary mapping or holography is well known in the context of black hole thermodynamics\cite{hooft2009dimensionalreductionquantumgravity,Susskind1995,Bousso2002} and the AdS-CFT correspondence\cite{Hubeny_2015}.
In condensed matter physics, dimensional reduction has been demonstrated in spin and orbital models\cite{Batista2005,Nussinov2015}. A distinct bulk-boundary correspondence principle operates in topological insulators\cite{Essin2011} and Weyl semi-metals\cite{Armitage2018}. 
Here, we provide a particularly simple example of a bulk-boundary mapping that may be realized in a large family of materials.

\section{Bonding considerations in \lowercase{d}$^3$ 1T-TMDs}

The d$^3$ TMDs are known to exhibit distortions, with atoms clustering into linked rhombus-shaped patterns. Many studies\cite{Burdett1984,Kertesz1984,Canadell1989,Whangbo1992,Fang1997,Rocquefelte2000,He2016,Choi2018} have discussed possible underlying mechanisms. We present an effective approach that takes into account the geometry of the 1T structure and the orientations of the valence orbitals.

As chalcogens are strongly electronegative, each chalcogen atom pulls in two electrons towards itself. In d$^3$ materials (e.g., ReS$_2$), this leaves the transition metal atom with three valence electrons. As the chalcogens are saturated, these electrons must be shared with neighbouring transition metal atoms. When a pair of electrons is shared locally between two metal atoms, this results in a valence bond. If electrons are shared across all metal atoms, the result is metallic bonding. In this article, we consider local valence bonds. This is consistent with the insulating (or semiconducting) behaviour of the materials under consideration. 

\begin{figure*}
\includegraphics[width=2\columnwidth,trim={0 16cm 0 0}]{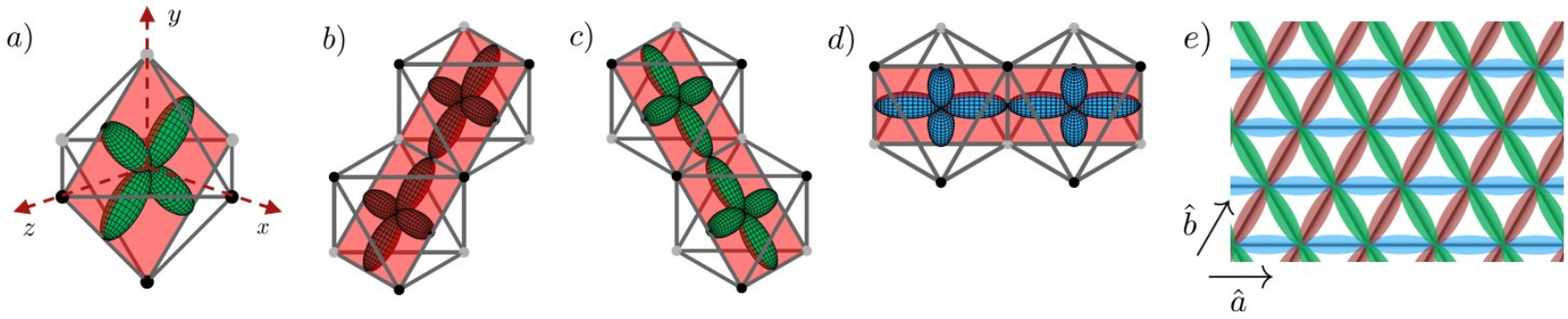}
\caption{a) Octahedral coordination in the 1T structure: the transition metal atom surrounded by six chalcogen atoms that are located at $\pm \hat{x}$, $\pm \hat{y}$ and $\pm \hat{z}$. The figure shows a $d_{yz}$ orbital, lying in the $yz$ plane. Overlaps of t$_{2g}$ orbitals on nearest neighbour bonds: b) $xy$ orbitals have strong overlaps along $\pm\{ \hat{x}+\hat{y}\}$ bonds, c) $yz$ orbitals along $\pm\{ \hat{y}+\hat{z}\}$ bonds and d) $zx$ orbitals along $\pm\{ \hat{x}-\hat{z}\}$ bonds. Each orbital exhibits strong overlap along a particular direction of the triangular lattice as shown in e), where the red, green, and blue dimers represent $xy$, $yz$ and $zx$ orbitals respectively.}
\label{fig.Orbitals}
\end{figure*}

In the 1T structure, each metal atom is octahedrally coordinated. The resulting crystal field splitting ensures that the three valence electrons reside in t$_{2g}$ orbitals. Remarkably, these t$_{2g}$ orbitals have strong directionality\cite{Burdett1984} as shown in Fig.~\ref{fig.Orbitals}. 
Taking a transition metal atom as the origin, we may define $\hat{x}$, $\hat{y}$ and $\hat{z}$ to point towards the vertices of the octahedral cage. This reference atom has six neighbouring transition metal atoms located at $\pm (\hat{x}+\hat{y})$, $\pm (\hat{y}+\hat{z})$ and $\pm (\hat{x}-\hat{z})$. We identify these vectors with the primitive translation vectors of the transition metal triangular lattice: $\hat{a} = \hat{x} - \hat{z}$, $\hat{b} = \hat{x}+\hat{y}$, with $\hat{b}-\hat{a}=\hat{y}+\hat{z}$. 
Along each nearest-neighbour bond of the triangular lattice, a certain orbital has maximal overlap. For example, along bonds in the $\pm \hat{a}$ direction, the $d_{zx}$ orbitals overlap strongly as they have lobes that point towards one another. At the same time, these bonds have zero overlap between (or even across) d$_{xy}$ and $d_{yz}$ orbitals due to symmetry. This orbital geometry imbues each valence bond with sharp orbital character: 
\begin{itemize}
    \item Bonds along $\hat{a}$ or $-\hat{a}$ are formed by sharing electrons that reside on the $d_{zx}$ orbitals of the respective metal atoms.
    \item Bonds along $\pm\hat{b}$ involve electrons on $d_{xy}$ orbitals.
    \item Bonds along $\pm (\hat{b}-\hat{a})$ 
    involve those on $d_{yz}$ orbitals.
\end{itemize}

Based on these ideas, we propose a model for valence bond formation in the following sections.

\section{Valence Bond Network Model: Configuration space}
\label{sec.configspace}

We assert that bonding configurations of d$^3$ TMDs can be mapped to networks using the following two rules:
\begin{itemize}[leftmargin=*]

\item Trivalent network rule: Dimers (valence bonds) are placed on nearest neighbour bonds of a triangular lattice (see Fig.~\ref{fig.Orbitals}), with three dimers touching each site. 
\end{itemize}
	
Each site of the lattice represents a metal atom, while each dimer represents a valence bond. With three dimers at each site, we naturally form a network by tracing connected dimers. As a result, each allowed configuration can be viewed as a `network covering' of the triangular lattice.
	
\begin{itemize}[leftmargin=*] 
\item Bending constraint: Two dimers that touch a site cannot be parallel. 
\end{itemize}

  \begin{figure}
		\includegraphics[width=\columnwidth,trim={0 16cm 5cm 0},clip]{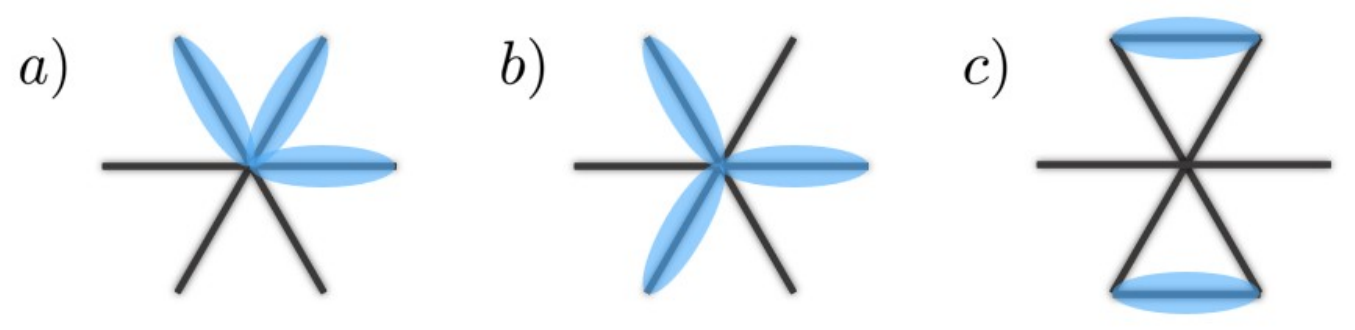}
		\caption{Elements of the model: Each site can host either (a) a `$\Psi$' motif or (b) a `$Y$' motif. The motifs may be rotated by any multiple of 60$^\circ$. (c) An hourglass motif formed by two parallel dimers. }
		\label{fig.PEshapes}
    \end{figure}
    
This rule arises from orbital structure. If one atom forms two parallel bonds (say along $\hat{a}$ and $-\hat{a}$ directions), it must have two electrons that reside on the same orbital ($d_{zx}$ in this case). This will impose an energy cost due to intra-orbital Coulomb repulsion. Assuming strong intra-orbital repulsion, we forbid such configurations via the bending constraint.

\begin{figure}
		\includegraphics[width=\columnwidth,trim={0 9cm 6.5cm 0},clip]{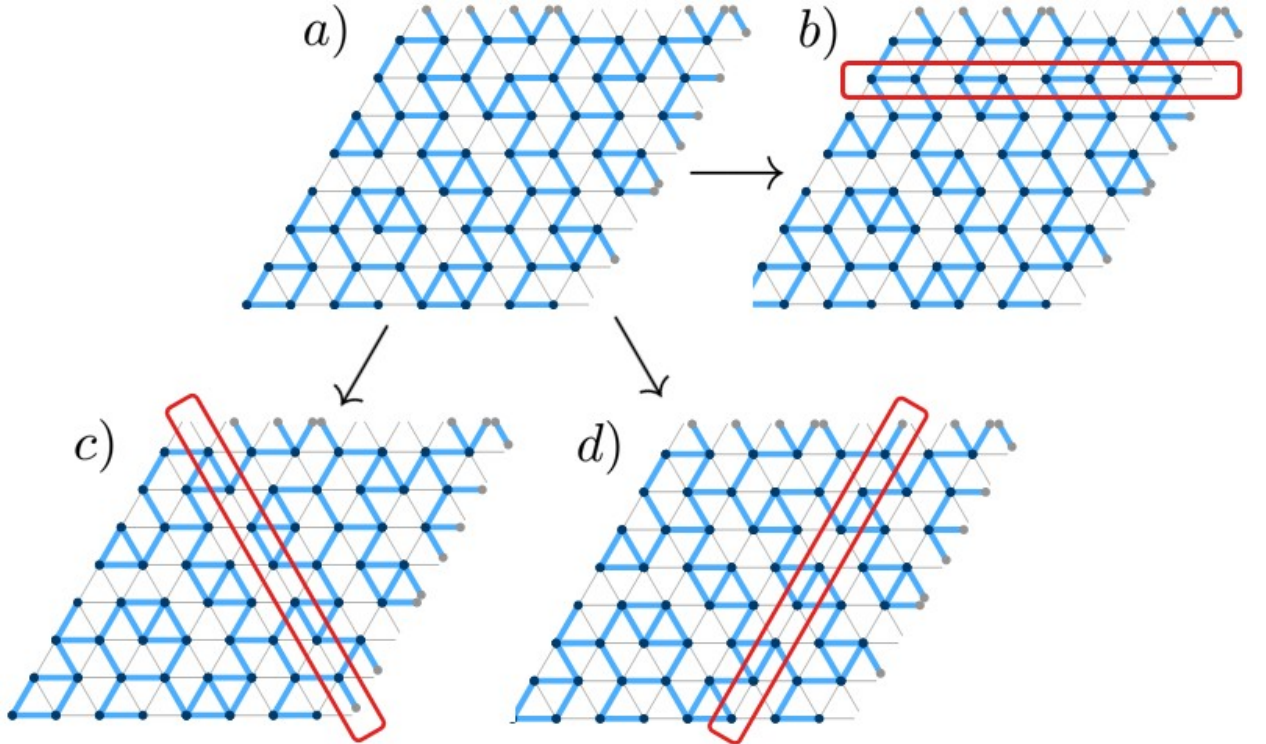}
		\caption{a) A dimer arrangement that satisfies the trivalent network rule and the bending constraint.
        Other valid arrangements (b, c, d) can be reached from configuration (a) by shifting bonds along a line of the underlying triangular lattice as shown.
	}
    \label{fig.ConfigShifts}
	\end{figure}

Remarkably, these two rules lead to a highly structured space of configurations. We list two corollaries:
\begin{enumerate}[leftmargin=*]
\item The three dimers that touch at a site must either form a `$\Psi$' (forming two adjacent acute angles, each of 60$^\circ$) or a `$Y$' (three adjacent obtuse angles of 120$^\circ$) as shown in Fig.~\ref{fig.PEshapes}(a,b). These are the only local configurations that satisfy the trivalent network rule \textit{and} the bending constraint. 
\item Any straight line of the triangular lattice (e.g., a line of bonds along $\hat{a}$) must have alternating dimers and blanks.
\end{enumerate}
The second corollary is a direct consequence of the first, as seen from the $\Psi$ and $Y$ configurations of Fig.~\ref{fig.PEshapes}(a,b). If a bond hosts a dimer, the adjacent parallel bond must necessarily be empty. Conversely, if a bond is empty, the adjacent parallel bond must host a dimer. An example of a bonding configuration is shown in Fig.~\ref{fig.ConfigShifts}(a).

With these considerations, we may enumerate all elements of the configuration space, i.e., all allowed dimer arrangements. Consider a triangular lattice with $L_a \times L_b$ sites and periodic boundary conditions, i.e., where translations along $L_a \hat{a}$  and $L_b \hat{b}$ are identified with the identity operation. Here, $L_a$ and $L_b$ represent the number of sites in the cluster along the $\hat{a}$ and $\hat{b}$ lattice directions. This system has $L_b$ lines that are parallel to $\hat{a}$ and $L_a$ lines that are parallel to $\hat{b}$. In addition, it has $L_c=\mathrm{hcf}(L_a,L_b)$ lines that are parallel to $(\hat{b}-\hat{a})$, where  $\mathrm{hcf}$ denotes the highest common factor; see Appendix~\ref{app.lines}. Each line must have alternating dimers and blanks. We may fix one particular bond either as hosting a dimer or as a blank. This immediately fixes all other bonds on the same line. We thus have a twofold choice on each line. These choices are independent and exhaustive -- they account for all possible configurations. We conclude that the number of allowed configurations is
\begin{equation}
N(L_a,L_b) = 2^{L_a + L_b + \mathrm{hcf}(L_a,L_b)}.
\label{eq.Nconfs}
\end{equation}

\section{Energetics}
We next consider energetics of bonding configurations that satisfy the rules described above. In the spirit of Rokhsar-Kivelson models\cite{Rokhsar1988}, kinetic energy can be ascribed to local rearrangements of dimers. In our system, no local rearrangement is possible. A change in one bond (dimer-to-blank or blank-to-dimer) immediately forces the adjacent parallel bonds to change. This effect propagates until we have a long-ranged rearrangement where all bonds along a line are changed, as shown in Fig.~\ref{fig.ConfigShifts}. Such long-ranged processes are unlikely to occur, especially when the system size is large. In this light, we neglect all kinetic energy terms.    

We consider potential energy contributions that arise from pairwise interactions between dimers. A pair of dimers may lie on parallel or non-parallel bonds. We argue that all pairwise interactions on non-parallel bonds can be folded into a single interaction parameter. This can be seen as follows. At each site, the three dimers that touch must either form a `$\Psi$' or a `$Y$' configuration as shown in Fig.~\ref{fig.PEshapes}(a,b). We associate an energy cost with each of these: $V_\Psi$ and $V_Y$.
At first glance, they appear to encode local Coulomb interactions between proximate dimers. However, when a $\Psi$ or a $Y$ motif is placed on the lattice, it immediately generates long-ranged correlations -- see Appendix~\ref{app.interdimer}. These impose energy costs arising from pairwise interactions over arbitrarily long distances. Crucially, all such non-trivial correlations are on non-parallel  bonds. 
It follows that all two-dimer interactions on non-parallel bonds can be subsumed into $V_\Psi$ and $V_Y$. In the rest of this article, we set $V_Y=0$ without loss of generality.

With pairwise interactions between dimers on parallel bonds, we only keep the shortest non-trivial contribution. This can be deduced as follows.
Consider a reference dimer on a given bond. Its presence immediately fixes dimers and blanks on all bonds that are located on the same line. As a result, we may neglect their interactions with the reference dimer -- they provide the same contribution in any valid configuration. We next consider dimers on a line that is immediately adjacent and parallel to that of the reference dimer. We have two possibilities on this line ($\ldots$-dimer-blank-$\ldots$ or $\ldots$-blank-dimer-$\ldots$). It can be easily seen that both possibilities place dimers at the same separations from the reference dimer. As a result, potential energy is the same for both possibilities -- a constant energy contribution that can be neglected. The shortest non-trivial interaction is with a parallel bond that is one line away, as shown in Fig.~\ref{fig.PEshapes}(c). If these bonds host dimers, they form a local `hourglass' configuration. We associate an energy cost $V_{\hourglass}$ with every hourglass. 

Despite the local character inherent in its definition, the hourglass term encodes a long-ranged interaction. This can be seen as follows. An hourglass configuration has two dimers on parallel bonds that are located one line apart. We now consider the lines on which these dimers reside. Due to the constraints on our configuration space, these lines will have alternating dimers and blanks. Consequently, if we have an hourglass configuration at one location, the immediately adjacent bonds will have blanks, the next pair will host dimers and so on. This will lead to a series of hourglass configurations repeating along the lines. The total energy contribution will be proportional to the linear system size (length of the line). This property magnifies the importance of the $V_{\hourglass}$ term.

With these considerations, the energy of any allowed configuration is $E_{\mathrm{config}}=N_{\Psi}V_{\Psi}+N_{\hourglass}V_{\hourglass}$, where $N_{\Psi}$ and $N_{\hourglass}$ count the number of $\Psi$ and hourglass motifs in the configuration. We may write this as 
	\begin{align}
		E_{\mathrm{config}} = &\sum_{i=1}^{N}\left[V_\Psi \sum_{j=1}^{6} 
		\big\vert  \begin{gathered}\includegraphics[width=0.2in]{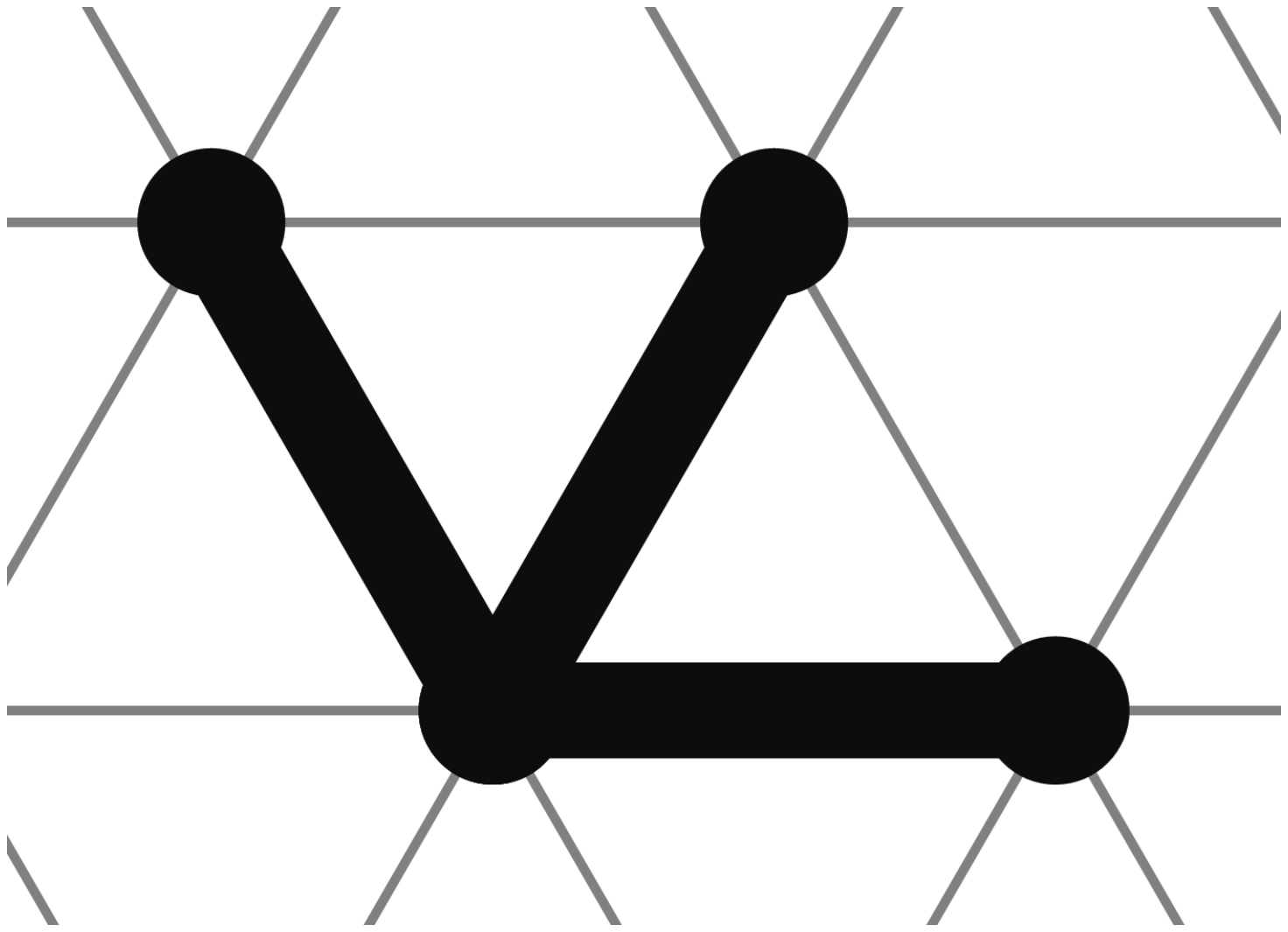} \end{gathered}
		\big\rangle \big\langle \begin{gathered} \includegraphics[width=0.2in]{Trident}\end{gathered} \big\vert 
		+V_{\hourglass} \sum_{k=1}^{3}\big\vert  \begin{gathered}\includegraphics[width=0.1in]{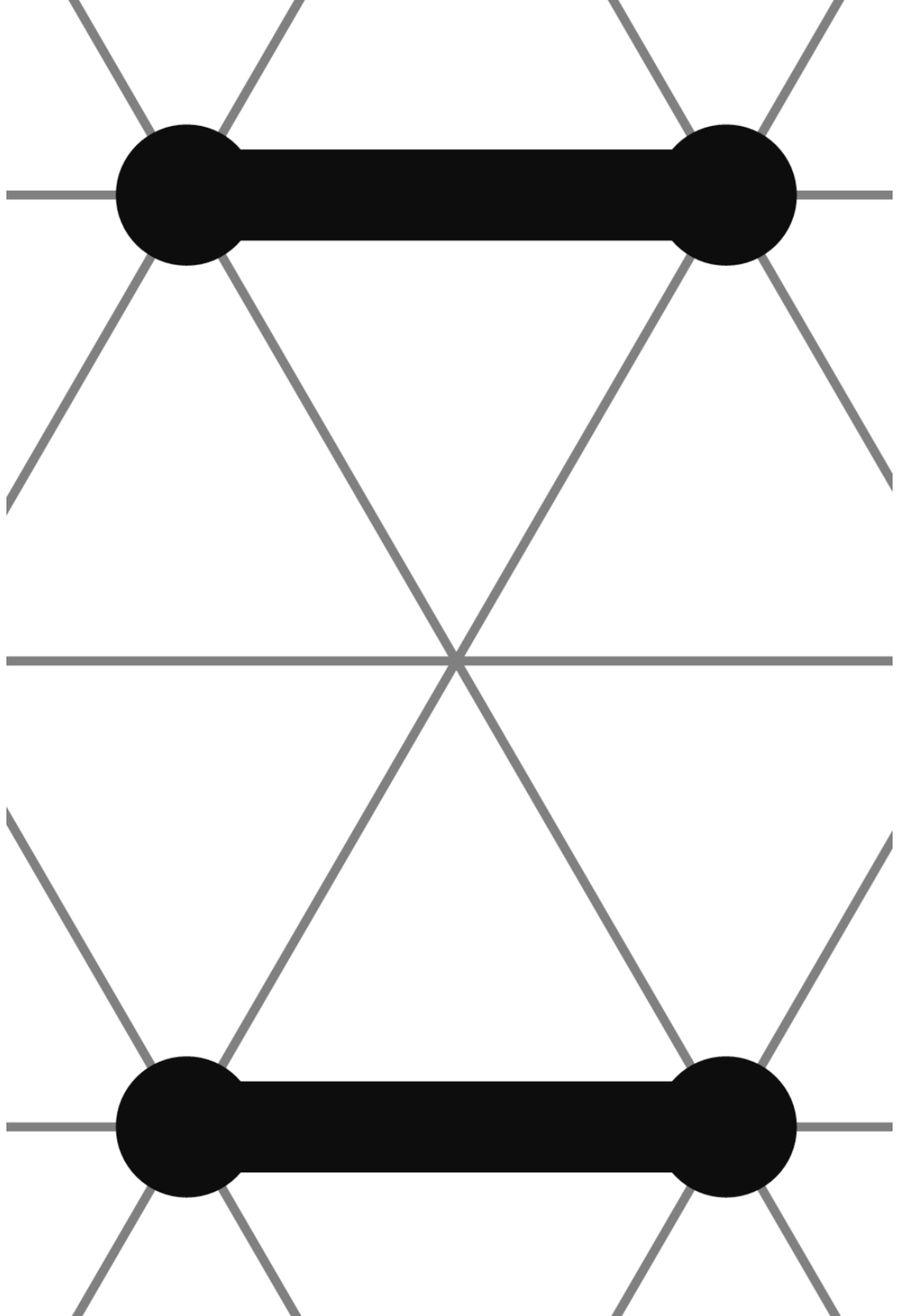} \end{gathered}
		\big\rangle \big\langle \begin{gathered} \includegraphics[width=0.1in]{Hourglass}\end{gathered} \big\vert 
		\right].
		\label{eq.Hamiltonian}
	\end{align}
	The sum over $i$ runs over all sites of the triangular lattice. The $V_\Psi$ term sums over six possible orientations of a $\Psi$ motif centered at site $i$. Similarly, the $V_{\hourglass}$ term involves a sum over three orientations of an hourglass attached to site $i$. The three orientations are chosen so as to avoid double counting when summed over $i$.

	\section{Numerical approach}
	To solve for the ground state of Eq.~\eqref{eq.Hamiltonian}, we first enumerate all valid configurations on an $L_a \times L_b$ cluster with periodic boundaries. As discussed in Sec.~\ref{sec.configspace}, this is equivalent to a two-fold choice on each of $L_a + L_b + \mathrm{hcf}(L_a,L_b)$ lines in the cluster. On each line, the two-fold choice corresponds to a (...-dimer-blank-dimer-...) or (...-blank-dimer-blank-...) arrangement. We algorithmically generate all such possibilities and calculate their energies according to Eq.~\eqref{eq.Hamiltonian}. We select configurations with minimum energy. After discounting degeneracies that arise from symmetry operations such as translations or reflections, we identify the pattern of dimers as a specific phase. We present a phase diagram as a function of $V_\Psi$ and $V_{\hourglass}$. 

    System size plays a crucial role in the numerical determination of ground states. We first note that $L_a$ and $L_b$ must necessarily be even numbers in order to host valid configurations. Otherwise, we will have lines with an odd number of bonds that cannot host an alternating sequence of dimers and blanks as required. Secondly, we require that $L_a$ and $L_b$ be multiples of four. This represents a commensurability condition, required for the true ground state(s) to fit within the considered system size. 
    Choosing $L_a$ and/or $L_b$ as a non-multiple of four leads to `frustration' with higher energy and a large system-size-dependent degeneracy. We discuss numerical results for frustrated cases in Appendix~\ref{app.frustrated}.

In the section below, we present a phase diagram deduced from numerical results for $(L_a,L_b)=(4,4), (4,8), (8,8), (4,12), (4,16)$ and $(8,12)$ -- system sizes accessible within our numerical resources. Note that the configuration space grows exponentially with system size as seen from Eq.~\eqref{eq.Nconfs}. We find precisely the same phase diagram for each of these $(L_a,L_b)$ values. We conjecture that this phase diagram will continue to hold as we approach the thermodynamic $L_a, ~L_b \rightarrow \infty$ limit.
    Each phase has a limited degeneracy that arises from symmetry operations such as translations, rotations, etc. This suggests that the true ground state is commensurate, i.e., it `fits' within the considered system dimensions.

        \begin{figure}
            \includegraphics[width=\columnwidth]{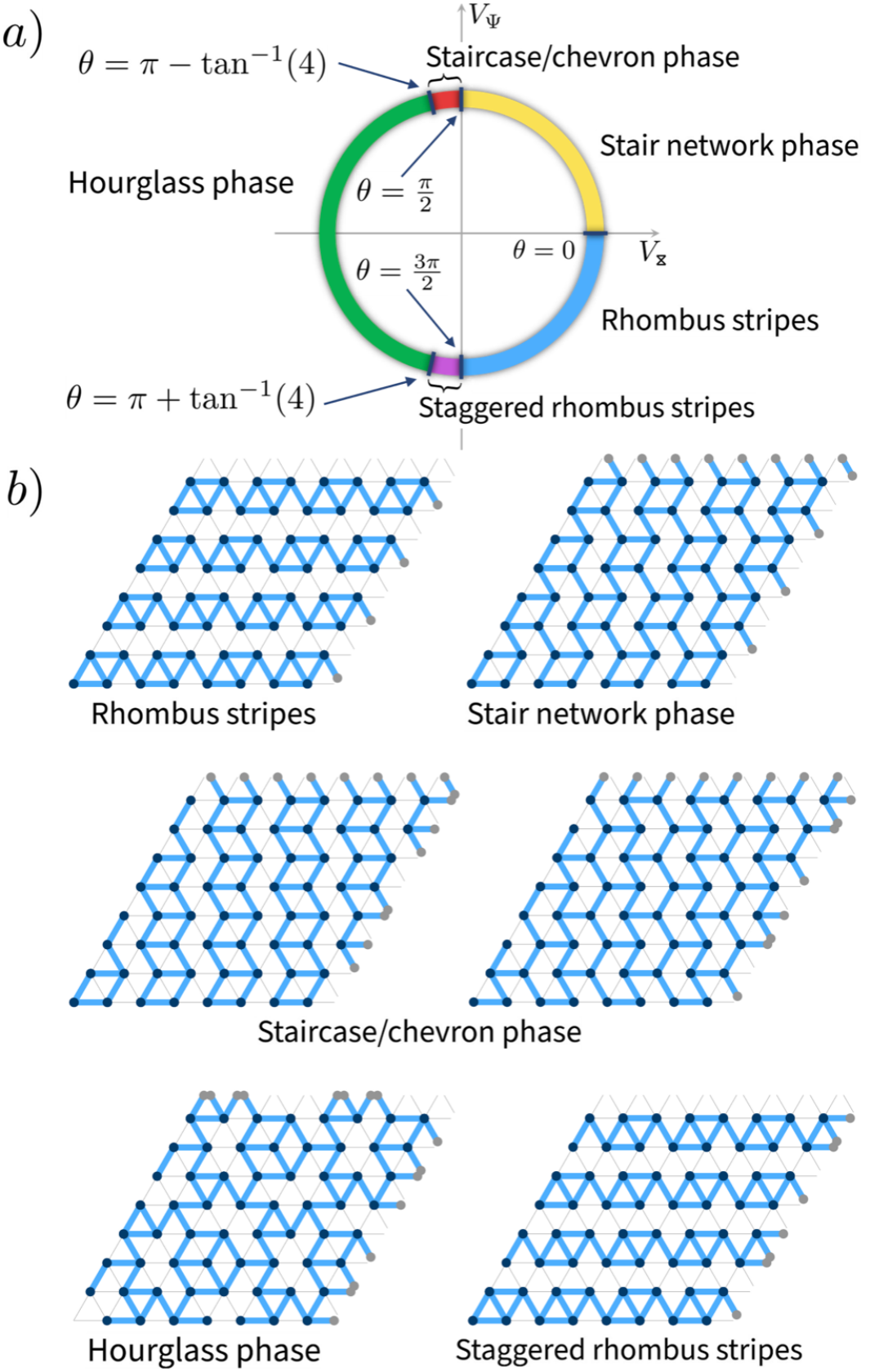}
    		\caption{$a)$ Phase diagram for the ground state in unfrustrated systems. $b)$ Representative configurations on an $(L_a,L_b)=(8,8)$ cluster for phases that appear in the phase diagram. For the Staircase/chevron phase, the staircase phase is shown on the left and the chevron phase on the right.}
    		\label{fig.PD_UF}
        \end{figure}

\section{Phase diagram}
\label{sec.phasediagram}

    We have two independent parameters, $V_\Psi$ and $V_{\hourglass}$. We can view these as a single degree of freedom, apart from an overall scaling of energy. 
    We define an angular variable $\theta$, where $V_\Psi \equiv \sin\theta$ and $V_{\hourglass} \equiv \cos\theta$. We show the phase diagram as a polar plot in Fig.~\ref{fig.PD_UF}, along with representative configurations.

   We find five phases:
    \begin{enumerate}[label=(\roman*),wide, labelwidth=!, labelindent=0pt]
    		\item \textbf{Rhombus stripe:} This phase appears for $3\pi/2 <\theta<2 \pi$. It has a $\Psi$ motif at every site with no hourglasses. This phase appears with a ground state degeneracy of 24 --- generated by four translations (including identity), three rotations (by 0$^\circ$, 120$^\circ$ and 240$^\circ$) and a mirror operation (with a two-fold degeneracy where each configuration is either left unchanged or is reflected). We identify this phase with the distortion seen in several d$^{3}$ materials, as described in Sec.~\ref{sec.materialdistort} below.
            \item \textbf{Stair network phase:} This phase appears for $0<\theta<\pi/2$. Although the coupling constants disfavour $\Psi$ motifs, they cannot be avoided entirely, i.e., it is impossible to tile the lattice using $Y$ motifs alone. This ground state has no hourglasses, consistent with the positive value of $V_{\hourglass}$.
            The ground state degeneracy is 24, generated by the same operations as the rhombus stripe. 
            \item \textbf{Staircase/chevron phase:} This phase appears for $\pi/2<\theta<\pi-\tan^{-1}(4)$. The arrangement of dimers either resembles a staircase or a chevron-tiling as shown in Fig.~\ref{fig.PD_UF}(b).  
            We have hourglasses along one direction of the triangular lattice. $Y$ motifs occur at half the sites. The ground state degeneracy is 24. This arises from three rotations and a mirror operation. In addition, we have four shift operations. In the direction perpendicular to the `staircase', we may shift dimers on all odd lines or all even lines independently.
            \item \textbf{Hourglass phase:} This phase appears for $\pi-\tan^{-1}(4)<\theta<\pi+\tan^{-1}(4)$. Hourglasses are maximized.
            There are 64 ground states. Along each of the three directions of the triangular lattice, we may shift dimers on all even lines or on all odd lines -- resulting in 4$^3$ independent shift operations. We discuss this degeneracy from a different perspective in Sec.~\ref{sec.effective} below. 
            
            \item \textbf{Staggered rhombus stripe:} 
            This phase appears for $\pi+\tan^{-1}(4)<\theta<3\pi/2$. 
            Hourglasses occur along one direction, with $\Psi$ motifs at every site.
            Starting from the `rhombus stripe' phase, this phase can be obtained by shifting every other stripe. The ground state is 24-fold degenerate, generated by the same operations that account for the degeneracy of the rhombus stripe. 
        \end{enumerate}
        
        All phase boundaries can be obtained analytically by comparing energies of phases on either side. To give an example, the staircase phase has energy $E_{staircase}=\frac{N}{2}\{ V_{\Psi}+V_{\hourglass}\}$, where $N$ is the number of sites. The hourglass phase has energy $E_{hourglass}=\frac{3N}{4}\{ V_{\psi}+2 V_{\hourglass}\}$. The transition between these phases occurs where $E_{staircase} = E_{hourglass}$, which corresponds to $V_\Psi/V_{\hourglass} = -4$ or $\theta= \pi-\tan^{-1} (4)$.

 \begin{figure}
    \includegraphics[width=\columnwidth]{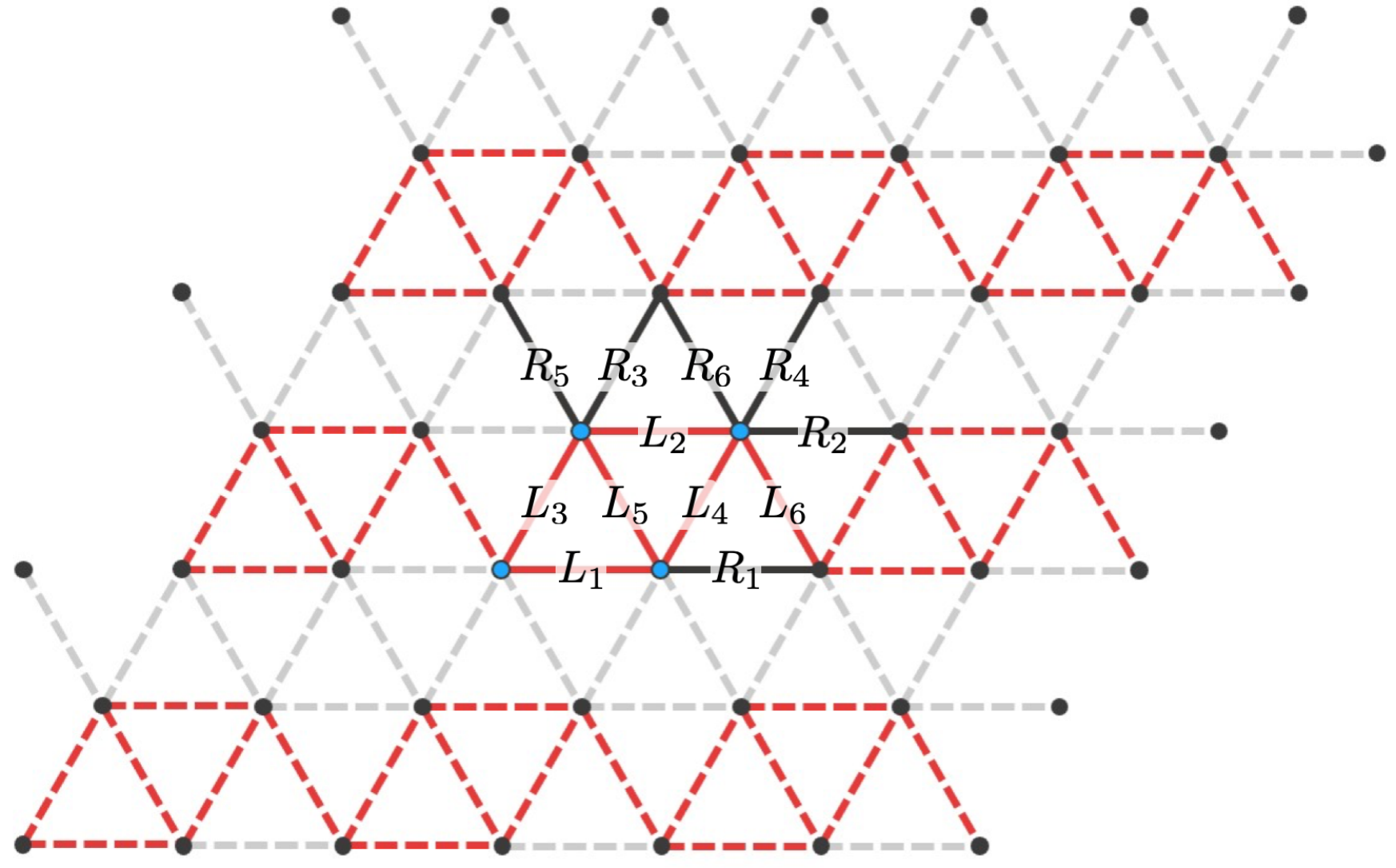}
		\caption{Distortion pattern seen in various d$^3$ materials. Shorter bonds are shown in red and labeled as $L_{1-6}$, while longer bonds are shown in black and labeled as $R_{1-6}$. Bonds within the unit cell are shown in solid lines, with bond lengths tabulated in Table~\ref{DistanceTable} for several materials. The  pattern shown directly correlates with the rhombus stripe phase.
        }
		\label{fig.interatomic}
	\end{figure}

    \begin{table*}
    	\begin{tabular}[t]{|l||l|l|l|l|l|l|l|l|}
    		\hline
    		Distance & ReSe$_{2}$\cite{Alcock1965} & ReSe$_{2}$\cite{Lamfers1996} & ReS$_{2}$\cite{Murray1994} & ReS$_{2}$\cite{Lamfers1996} &ReSSe\cite{Lamfers1996}& TcS$_{2}$\cite{Lamfers1996} &NaMoO$_{2}$\cite{Vitoux2020}
            &LiMoS$_2$\cite{SchwarzmüllerStefan2024PSoH}\\
            \hline
            \hline
    		$L_{1}$ & 2.92 & 2.8714 & -     & 2.805 & 2.866 & 2.8156 & 2.688 & 2.9847 \\
    		\hline
            $L_{2}$ & 2.92 & 2.8714 & 2.824 & 2.801 & 2.866 & 2.8156 & 2.688 & 2.9847 \\
    		\hline
    		$L_{3}$ & 2.83 & 2.8443 & -     & 2.800 & 2.805 & 2.7883 & 2.648 & 2.9110 \\
    		\hline
            $L_{4}$ & 2.83 & 2.8443 & 2.790 & 2.764 & 2.805 & 2.7883 & 2.648 & 2.9110 \\
    		\hline
            $L_{5}$ & 2.64 & 2.7333 & 2.695 & 2.693 & 2.694 & 2.6997 & 2.577 & 2.8024 \\
    		\hline
    		$L_{6}$ & 3.07 & 2.9867 & 2.895 & 2.888 & 2.944 & 2.8981 & 2.704 & 3.0459 \\
    		\hline
    		$R_{1}$ & 3.68 & -      & -     & -     & -     & -      & -  & -   \\
    		\hline
            $R_{2}$ & 3.68 & -      & 3.560 & -     & -     & -      & -  & -   \\
    		\hline
            $R_{3}$ & 3.91 & -      & -     & -     & -     & -      & -   & -  \\
    		\hline
            $R_{4}$ & 3.91 & -      & 3.744 & -     & -     &        & -   & -  \\
    		\hline
    		$R_{5}$ & 4.12 & -      & -     & -     & -     & -      & -   & -  \\
    		\hline
    		$R_{6}$ & 3.81 & -      & -     & -     & -     & -      & -  & -   \\
    		\hline
    	\end{tabular}
    	\caption{Interatomic distances between transition metal atoms within an MX$_{2}$ layer in several materials. Distances are given in units of Angstroms, based on the labels in Fig.~\ref{fig.interatomic}. In all cases, bonds $L_{1-6}$ are shorter than $R_{1-6}$. Entries show bond distances reported in literature, with blanks where data is not available. }
    	\label{DistanceTable}
    \end{table*}

    \section{Rhombus stripes in 1T TMDs}
    \label{sec.materialdistort}
        Several d$^3$ TMD's exhibit a distorted 1T structure that  is characterized by linked rhombus-like clusters. The unit cell for this distorted structure is shown in Fig.~\ref{fig.interatomic}. The figure only shows the transition metal atoms, disregarding the positions of the chalcogens. The unit cell contains six shorter bonds (labeled as $L_{1-6}$) and six longer bonds (labeled as $R_{1-6}$). The materials that exhibit this distortion are listed as columns in Table~\ref{DistanceTable}. Bond lengths are tabulated for each material, based on published data.

        We assert that this distortion is a manifestation of the rhombus-stripe phase in our phase diagram. This can be seen by comparing the unit cell of Fig.~\ref{fig.interatomic} with the rhombus-stripe phase ($3\pi/2 < \theta<2\pi$) in Fig.~\ref{fig.PD_UF}.
        In our model, a dimer indicates the presence of a valence bond which will, in turn, lead to a shorter inter-atomic distance. This offers a unified explanation for the distortion observed across several materials.

	\section{Impurity textures}
    \label{sec.impurities}
\begin{figure*} \includegraphics[width=2\columnwidth,trim={0 15cm 0 0}]{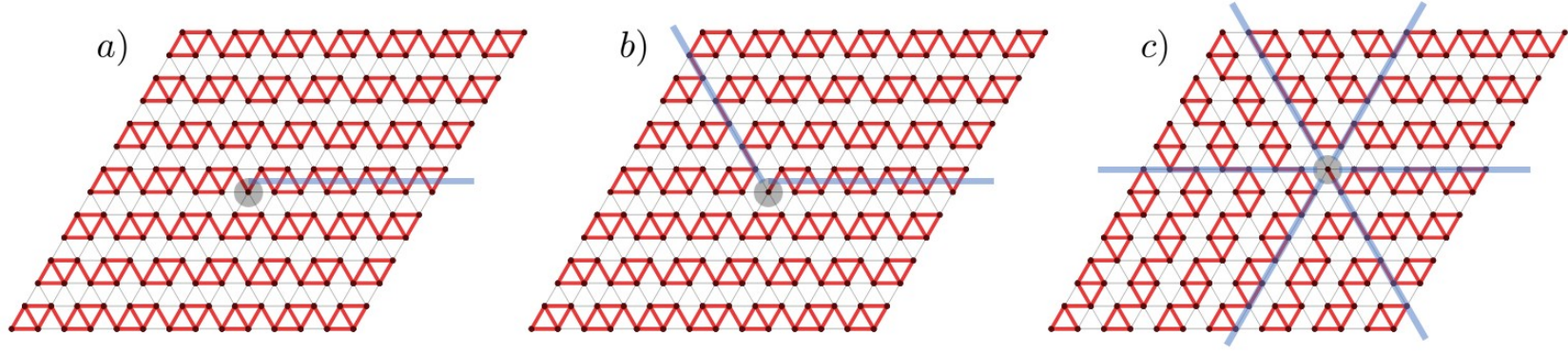}
		\caption{$a)$ Texture generated by a single d$^2$ impurity. $b)$ and $c)$ show textures generated by single d$^1$ impurities. Each configuration maximizes rhombus stripes and shows the neighbourhood of an impurity within a larger system.}
		\label{fig.impurities}
	\end{figure*}

    As a smoking-gun test for the validity of our model, we argue that isolated impurities will produce long-ranged textures. We consider substitutional impurities in the form of d$^1$ or d$^2$ atoms that sit at a d$^3$ transition metal site. For example, in ReS$_2$, we may have a Ta or W atom at a Re site. Within our model, this forces a local change to the constraints -- at the impurity site, we must have one (in the case of d$^1$) or two (in the case of d$^2$) dimers, instead of three. We can no longer have long-ranged patterns corresponding to any of the phases discussed above. Rather, we will have one or more domain walls that emanate from the impurity. This is illustrated in Fig.~\ref{fig.impurities}, depicting three ways by which a d$^1$ (or d$^2$) impurity may be accommodated in a rhombus-stripe phase. In each case, one or more domain walls stretches out from the impurity. Such textures can be mapped by scanning probes on layered TMDs. 

\section{Effective boundary description}
\label{sec.effective}

    \begin{figure}
		\includegraphics[width=\columnwidth,trim={0 8cm 1cm 0}]{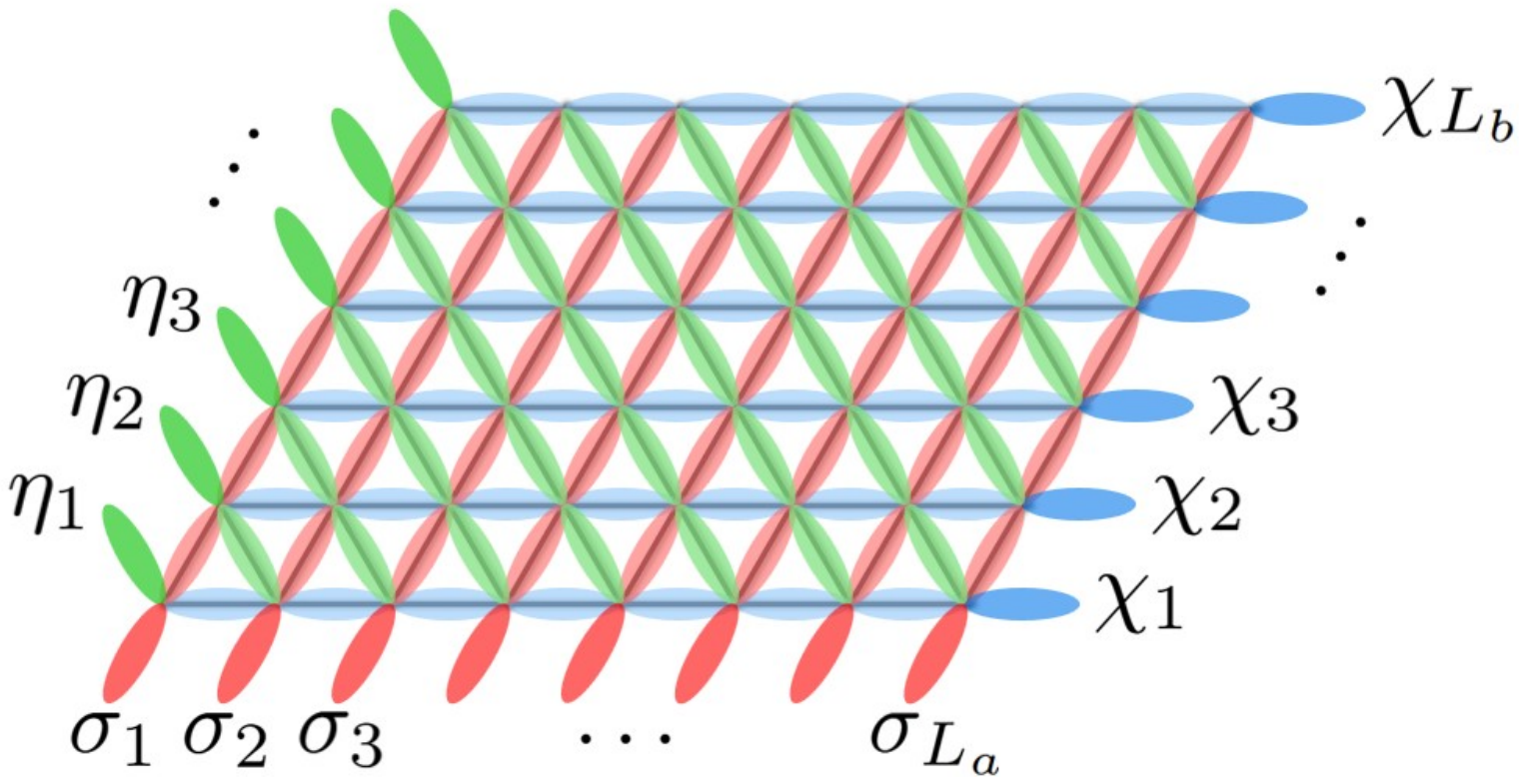}
		\caption{Ising variables along the boundary of an $L_a$ by $L_b$ lattice. The number of distinct $\eta$ variables is $L_c = \mathrm{hcf}(L_a,L_b)$.}
		\label{fig.I_G}
    \end{figure}
    
The degrees of freedom in our model reside in the bulk of the triangular lattice, with each bond either hosting a dimer or a blank. In a system with open boundaries, the bulk configuration is entirely determined by the dimer configuration at the boundaries. This is illustrated in Fig.~\ref{fig.I_G}, where we highlight three lines of bonds (parallel to the $\hat{a}$, $\hat{b}$ and $\hat{b}-\hat{a}$ directions) at the boundary. We define an Ising variable on each of these bonds, with $+1$ to represent a dimer and $-1$ to represent a blank. We arrive at a representation for the boundary configuration in terms of three Ising chains, denoted as $\{\sigma_i, \chi_j, \eta_k\}$, where ($i=1,\ldots, L_a$), $(j=1,\ldots,L_b)$ and $(k=1,\ldots,\mathrm{hcf}(L_a,L_b))$. Here, $\mathrm{hcf}(L_a,L_b)$ denotes the highest common factor of $L_a$ and $L_b$.

We first assert that the bulk configuration is entirely determined by the boundary. This follows from the property that each line must have an alternating sequence of dimers and blanks, as discussed in Sec.~\ref{sec.configspace}. When a bond on the boundary is fixed (as a dimer or a blank), this immediately fixes a line of bonds that runs through the bulk. As seen from Fig.~\ref{fig.I_G}, when bonds on the three boundary Ising chains are fixed, this fixes all bonds within the bulk.   
    
We next seek to express the bulk energy as given by Eq.~\eqref{eq.Hamiltonian} in terms of the boundary Ising variables. The $V_{\hourglass}$ term in Eq.~\eqref{eq.Hamiltonian} takes a particularly simple form -- as a next-neighbour coupling within each Ising chain. To see this, consider $\sigma_1$ and $\sigma_3$. If these Ising variables are opposite in value, we have parallel dimer arrangements on the lines along which the bonds lie. This leads to hourglasses being formed at every alternate bond along the lines. As the total number of bonds determined by $\sigma_1$ (or $\sigma_3$) is $L_b$, the number of hourglasses formed is $\frac{L_b}{2}$. If $\sigma_1$ and $\sigma_3$ carry the same value, no hourglasses are formed along these lines. This allows us to write the number of hourglasses between $\sigma_1$ and $\sigma_3$ lines as $L_b(1-\sigma_1 \sigma_3)/4$. Considering each pair of neighbouring lines in the same manner, we write    
\begin{eqnarray}
    \nonumber N_{\hourglass} = C&-&\frac{L_b}{4}\sum_{i=1}^{L_a} \sigma_i \sigma_{i+2} - \frac{L_a}{4}\sum_{j=1}^{L_b} \chi_j \chi_{j+2} \\
        &-& \frac{L_aL_b}{4\;\mathrm{hcf}(L_a,L_b)}\sum_{k=1}^{\mathrm{hcf}(L_a,L_b)} \eta_k \eta_{k+2},
        \label{eq.Nhourglass}
\end{eqnarray}
where $C= 3L_aL_b/4$.

The $V_\Psi$ term in Eq.~\eqref{eq.Hamiltonian} couples to the number of $\Psi$ motifs in the system. This quantity can be expressed in terms of the boundary Ising variables as
\begin{equation}
N_\Psi = \frac{1}{4}\left[3 L_aL_b + \frac{1}{L_c} (L_b M_\sigma A_\eta + L_a M_\chi M_\eta) + A_\sigma A_\chi\right],
\label{eq.NPsi}
\end{equation}
where $L_c=\mathrm{hcf}(L_a,L_b)$ and we have defined `magnetisation' variables
\begin{eqnarray}
M_\sigma = \sum_{i=1}^{L_a}  \sigma_i; ~~M_\chi = \sum_{j=1}^{L_b}  \chi_j; ~~ M_\eta = \sum_{k=1}^{L_c}  \eta_k, \label{eq.MDefs}
\end{eqnarray}    
and `staggered magnetisation' variables
\begin{equation}
A_\sigma = \sum_{i=1}^{L_a} (-1)^i \sigma_i; ~A_\chi = \sum_{j=1}^{L_b} (-1)^j  \chi_j;  ~A_\eta = \sum_{k=1}^{L_c}  (-1)^k\eta_k. \label{eq.ADefs}
\end{equation}    
A detailed derivation of Eq.~\eqref{eq.NPsi} is provided in Appendix~\ref{app.Ising}. Notably, $N_\Psi$ represents long-ranged couplings across the three Ising chains. This can be seen from the fact that $N_\Psi$ is written in terms of `macroscopic' magnetisation and staggered-magnetisation variables. 

We finally rewrite Eq.~\eqref{eq.Hamiltonian} as 
\begin{eqnarray}
E(\{\sigma \},\{ \chi\},\{\eta\}) = V_{\hourglass} N_{\hourglass} + V_\Psi N_\Psi, \label{eq.E_Ising}    
\end{eqnarray}
where $N_{\hourglass}$ and $N_\Psi$ are defined in Eqs.~\eqref{eq.Nhourglass} and \eqref{eq.NPsi} respectively -- both in terms of boundary variables. 
This boundary theory is consistent with the phase diagram shown in Fig.~\ref{fig.PD_UF}. As a concrete example, we explain the occurrence and parameter range of the hourglass phase. This phase appears when $V_{\hourglass}$ is negative and dominant. In order to maximize hourglasses, each Ising chain must have a configuration with a repeating pattern of $\{+1,+1,-1,-1,\}$. This maximizes the contribution to $N_{\hourglass}$ in Eq.~\eqref{eq.Nhourglass} above (i.e., it maximizes $(-)\sum_i \sigma_i \sigma_{i+2}$ and its analogues). This configuration fixes the magnetisation ($M$'s) and staggered magnetisation ($A$'s) variables to be zero. As a result, $N_\Psi$ is pinned to a constant value of $3L_aL_b/4$. A significant strength of $V_\Psi$ is required to induce a phase transition. See Appendix~\ref{app.IsingPhases} for descriptions of the other phases from Sec.~\ref{sec.phasediagram} in terms of the Ising chains.

\section{Discussion}
The trivalent network model provides an interesting example of locally constrained physics. In the context of 1T-TMDs, the model comes with an additional bending constraint. This immediately leads to a highly structured space of bonding configurations. This can be contrasted with polyacytelene -- a paradigmatic valence-bond system that allows for two possible bonding configurations. Here, in the 1T-d$^3$-TMDs, we have an exponentially large space of valence bond configurations. This allows for competition among a large number of bonding arrangements and a rich phase diagram.

Our results share many similarities with fractonic models\cite{Slagle2017,Nandkishore2019,Gromov2024}. The space of bonding configurations grows exponentially with linear system size. Dynamics is forbidden.
An exciting future direction is to introduce defects and to explore their role in dynamics. This may pave the way for fracton-like physics in materials. This can be achieved, for example, by varying Li concentration in LiMoS$_2$. This will lead to electron-deficient sites that act as defects. Unlike substitutional impurities shown in Fig.~\ref{fig.impurities}, these defects may not be pinned to specific sites. As argued in Sec.~\ref{sec.impurities}, a single defect will induce a long-ranged domain wall. With multiple defects, we may obtain confinement of pairs -- with two defects at the ends of a domain wall. Such a pair of defects may exhibit directional mobility, e.g., along the rhombus stripe direction. A large density of defects may lead to new distortion patterns\cite{Sun2018}, possibly arising from defect-condensation.

Our model offers an explanation for the observed rhombus-stripe distortion in ReS$_2$, ReSe$_2$, NaMoO$_2$, etc.
Our phase diagram hosts several other phases which suggest competing distortion patterns in these materials. They could potentially be induced by varying constituents or by applying pressure/ strain. We have demonstrated that the number of allowed valence bond configurations is exponentially large. This suggests the possibility of competition among a large number of bonding configurations, e.g., if the rhombus stripe phase were destabilized by pressure. Further possibilities include the application of electric fields to induce dynamical effects, where the material goes through a sequence of bonding configurations.

    \acknowledgments
        We thank Kirill Samokhin and Han Yan for useful discussions. This work was supported by the Natural Sciences and Engineering Research Council of Canada through Discovery Grant 2022-05240.
	
	\bibliography{main}

    \clearpage
    \appendix

    \setcounter{figure}{0}
    \setcounter{equation}{0}
    \renewcommand{\theequation}{A\arabic{equation}}
    \renewcommand{\thefigure}{A\arabic{figure}}
    
    \section{Lines on the triangular lattice}
    \label{app.lines}
\begin{figure}
            \includegraphics[width=.80\columnwidth]{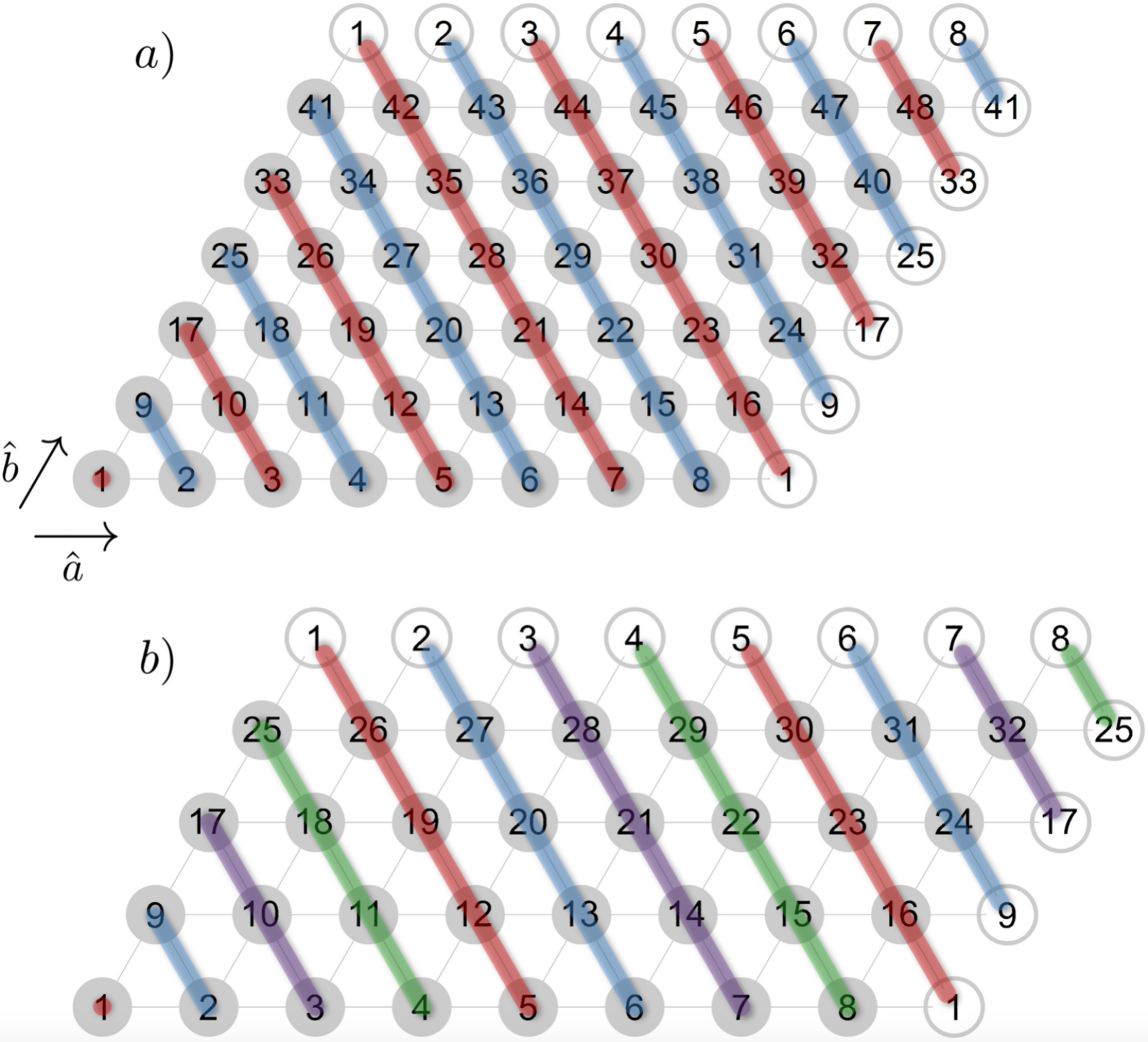}
    		\caption{Lines along the $\hat{b}-\hat{a}$ direction in triangular clusters with periodic boundaries. (a) With $(L_a,L_b)=(8,6)$, we have two distinct lines. (b) With $(L_a,L_b)=(8,4)$, we have four distinct lines. In each cluster, distinct lines are distinguished by colour.}
    		\label{fig.HCF}
        \end{figure}

    We find the number of distinct lines along the $\hat{b}-\hat{a}$ direction in a triangular cluster with periodic boundaries. We trace a closed line by starting at a site and moving in the $\hat{b}-\hat{a}$ direction until we return to the same site. Examples are shown in Fig.~\ref{fig.HCF} with lines wrapping around multiple times due to periodic boundary conditions.

    We consider a generic initial site located at $m\hat{a}+n\hat{b}$, where $m$ and $n$ are integers. After moving $j$ steps in the $\hat{b}-\hat{a}$ direction, we reach $(m-j)\hat{a}+(n+j)\hat{b}$. For this to be the initial site, the total displacement must be an integer combination of $L_a \hat{a}$ and $L_b \hat{b}$. We must have
    $j=\mu L_a = \nu L_b$, where $\mu$ and $\nu$ are integers. The smallest value of $j$ where this can hold true is the lowest common multiple of $L_a$ and $L_b$: $j=\text{lcm}(L_a,L_b)$. This yields the length of each line in the $\hat{b}-\hat{a}$ direction. 

    The number of distinct lines is obtained by dividing the total number of sites by the length of each line,
    \begin{equation*}
        L_c=\frac{L_a L_b}{j}=\frac{L_a L_b}{\text{lcm}(L_a,L_b)}=\text{hcf}(L_a,L_b).
    \end{equation*}

    \setcounter{figure}{0}
    \setcounter{equation}{0}
    \renewcommand{\theequation}{B\arabic{equation}}
    \renewcommand{\thefigure}{B\arabic{figure}}

    \section{Inter-dimer interactions}
    \label{app.interdimer}
\begin{figure}
        \includegraphics[width=\columnwidth,trim={0 13.5cm 0 -1cm}]{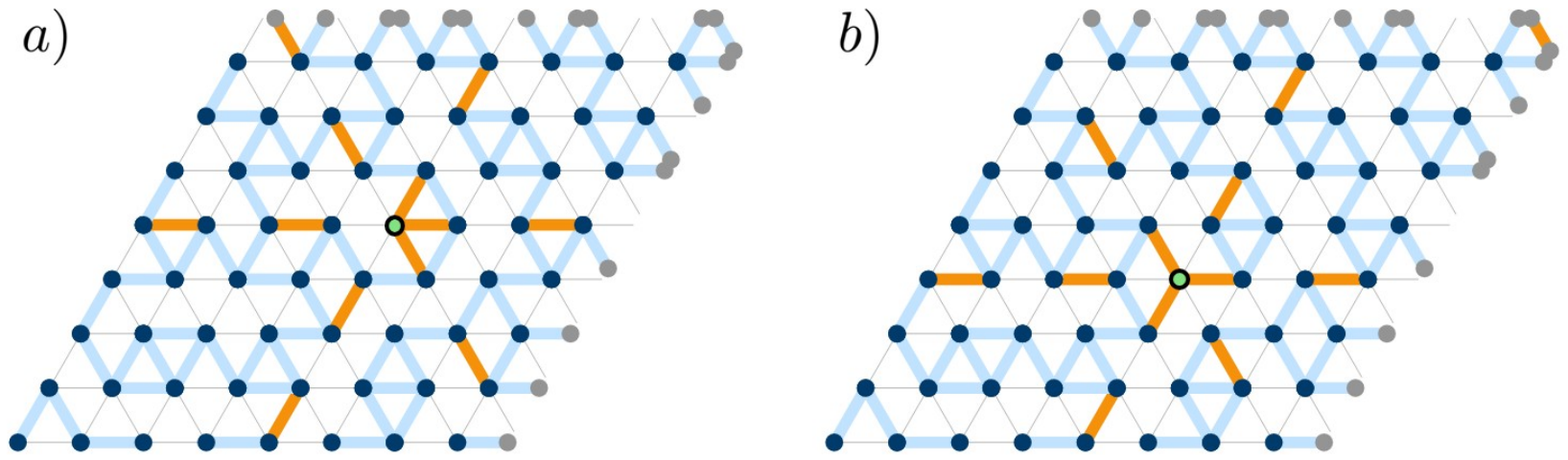}
        \caption{(a) Placing a $\Psi$ motif at a certain site immediately forces us to place bonds (shown in orange) at large distances from the site. The energy cost of a $\Psi$ motif includes correlations between all pairs among these dimers. (b) A $Y$ motif similarly induces dimers at long distances. }
        \label{fig.InterDim}
    \end{figure}

    As argued in Sec.~\ref{sec.configspace}, each site of the triangular lattice must either have a $\Psi$ or a $Y$ motif, with energy $V_\Psi$ or $V_Y$ respectively. We now argue that these energies are very general -- they include dimer-dimer interactions on any pair of non-parallel bonds. To see this, we compare placing a $\Psi$ motif vs.~a $Y$ motif at a certain site. In each case, we are immediately forced to place bonds at arbitrarily large distances from the site -- see Fig.~\ref{fig.InterDim}. This is due to the constraint that each line must have an alternating sequence of dimers and bonds. A first-principles estimate of $V_\Psi$ or $V_Y$ must include interaction costs between all pairs of dimers thus formed.

    A pair of dimers on any two non-parallel bonds can be traced to a $\Psi$ or $Y$ motif. This motif lies at the intersection of the lines along which the bonds lie. We conclude that the energy cost  associated with dimer-dimer interactions on any pair of non-parallel bonds is included in $V_\Psi$ and $V_Y$.
    
    \setcounter{figure}{0}
    \setcounter{equation}{0}
    \renewcommand{\theequation}{C\arabic{equation}}
    \renewcommand{\thefigure}{C\arabic{figure}}
    
    \section{Frustration and system size}
    \label{app.frustrated}

    When the system dimensions ($L_a$ and/or $L_b$) are not multiples of four, we argue that the true ground state does not fit within. We present three pieces of evidence to support this assertion.

\begin{figure}[!htb]
            \includegraphics[width=\columnwidth,trim={0 0 0 0}]{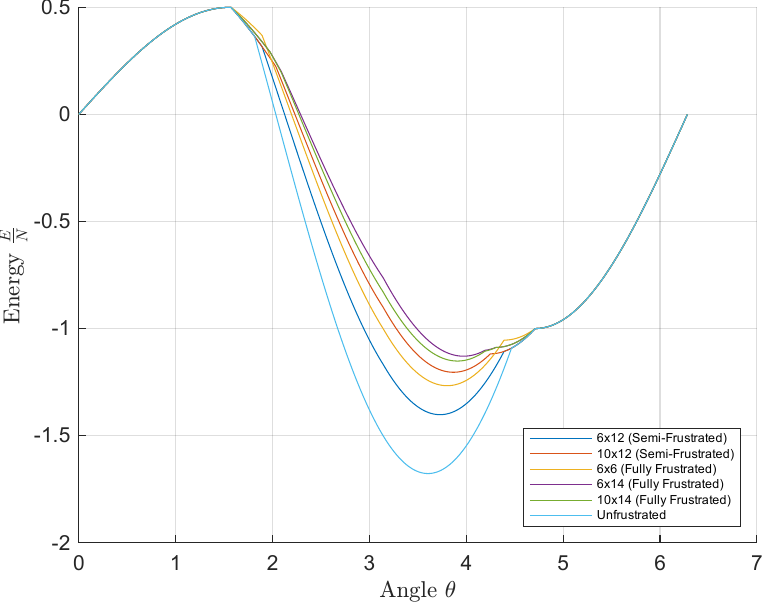}
    		\caption{Ground state energy (per site) as a function of $\theta$ for various system sizes. The `unfrustrated' case corresponds to $(L_a,L_b)=(4,4)$, $(4,8)$, $(4,12)$, $(4,16)$, $(8,8)$ and $(8,12)$ -- all of which produce the same curve.  }
    		\label{fig.EnergyPlot}
        \end{figure}

    First, we consider the ground state energy. Fig.~\ref{fig.EnergyPlot} plots ground state energy per site as a function of $\theta$ (as defined in Sec.~\ref{sec.phasediagram}) for various system sizes. The energy is lowest for `unfrustrated' system sizes (i.e., where $L_a$ and $L_b$ are multiples of four). Energy is either the same or higher for `semi-frustrated' (when $L_a$ is a multiple of four while $L_b$ is not, or the other way around) and `fully frustrated' (when both $L_a$ and $L_b$ are non-multiples of four) systems. Crucially, in the latter two cases, energy per site varies with system size -- unlike unfrustrated systems. This suggests that unfrustrated systems have the same dimer arrangement regardless of system size -- plausibly extending all the way to the thermodynamic limit.

\begin{figure}[!htb]
		\includegraphics[width=\columnwidth]{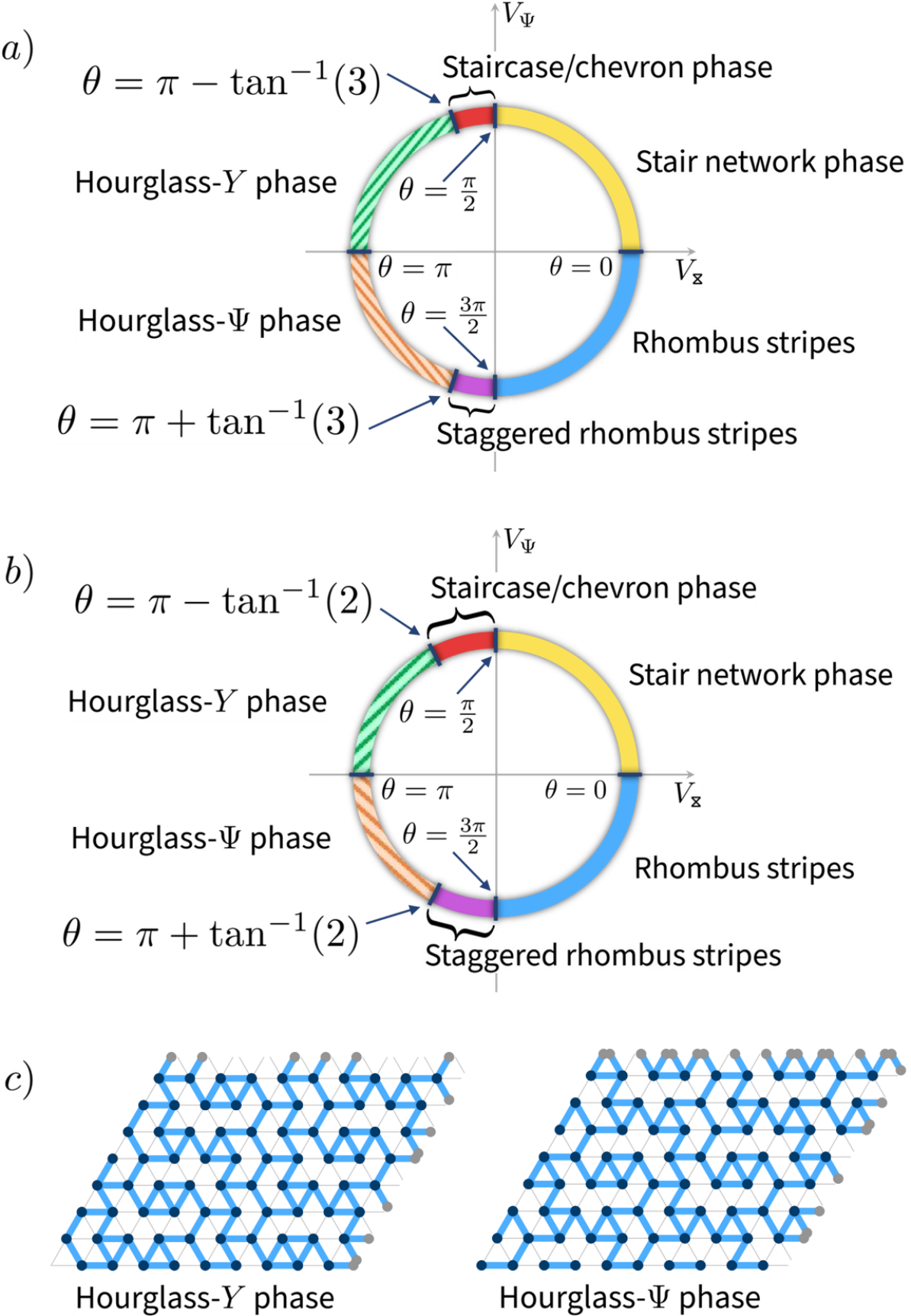}
		\caption{$a)$ Ground state phase diagrams for semi-frustrated systems. $a)$ $(L_a,L_b)=(6,12)$. $b)$ $(L_a,L_b)=(10,8)$. $c)$ Examples of configurations from the phase diagram shown in $(b)$.}
		\label{fig.PD_SF6x12}
    \end{figure}

\begin{figure}
		\includegraphics[width=\columnwidth]{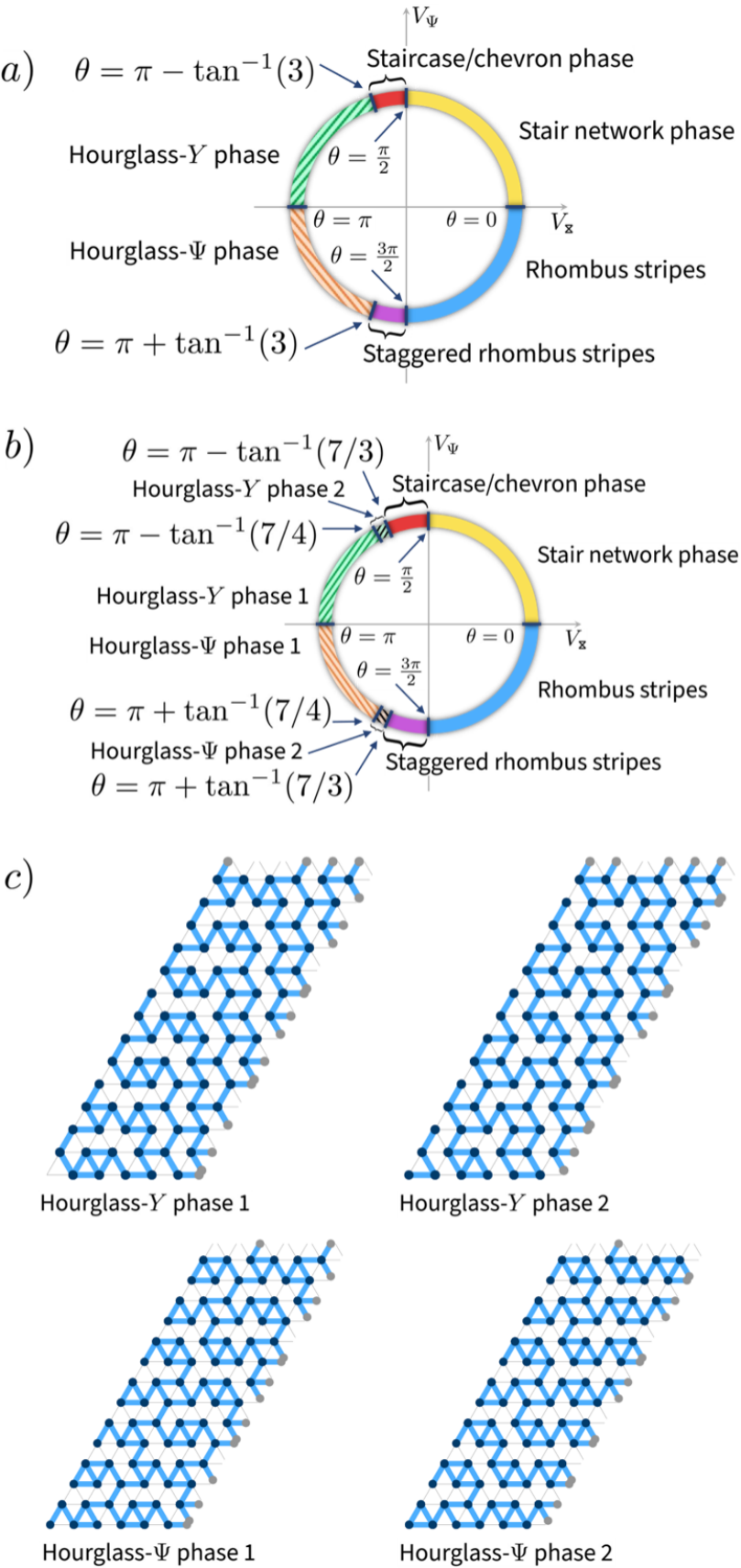}
		\caption{Phase diagrams for the ground state in fully frustrated systems. $a)$ $(L_a,L_b)=(6,6)$. $b)$ $(L_a,L_b)=(6,14)$. $c)$ Examples of configurations from the phase diagram in $(b)$. Hourglass-$Y$ phase 1 has the same number of hourglasses as Hourglass-$\Psi$ phase 1, but fewer $\Psi$ motifs. Hourglass-$Y$ phase 2 has fewer hourglasses and $\Psi$ motifs than Hourglass-$Y$ phase 1. Hourglass-$\Psi$ phase 2 has fewer hourglasses than Hourglass-$\Psi$ phase 1, but more $\Psi$ motifs. The staggered rhombus stripe and staircase/chevron phases are similar to those in unfrustrated systems, but with defects due to frustration.}
		\label{fig.PD_FF6x14}
    \end{figure}

    We next consider the ground state phase diagram. For all unfrustrated system sizes considered, we find the same phase diagram -- shown in Fig.~\ref{fig.PD_UF}. The phases and phase boundaries remain the same as size is varied. In contrast, frustrated systems show variations with system size. This can be seen from Figs.~\ref{fig.PD_SF6x12} (semi-frustrated) and \ref{fig.PD_FF6x14} (fully frustrated). We find variations in the nature of each phase as well as in the locations of phase boundaries.

    Finally, we consider the degeneracies of the lowest energy configuration. In unfrustrated cases, each phase in the phase diagram has a relatively small degeneracy that is fully accounted for by symmetry operations. This holds true for all system sizes explored. In contrast, frustrated cases show strong variations in degeneracy as system size is tuned. For example, an Hourglass-$Y$ phase appears in semi-frustrated cases, as shown in Fig.~\ref{fig.PD_SF6x12}. The degeneracy of this phase varies from 648 for $(L_a,L_b)=(6,12)$ to 200 for $(10,8)$. This large degeneracy cannot be attributed to any symmetry operations. Rather, it indicates an `accidental' degeneracy that arises from competition among a large number of configurations. This is strongly reminiscent of extended degeneracy seen in models of frustrated magnetism\cite{lacroix2011introduction}.


    \setcounter{figure}{0}
    \setcounter{equation}{0}
    \renewcommand{\theequation}{D\arabic{equation}}
    \renewcommand{\thefigure}{D\arabic{figure}}

    \section{Energy in the effective boundary theory}
    \label{app.Ising}
\begin{figure*}[h]
        \includegraphics[width=2\columnwidth,trim={0 12cm 0 0}]{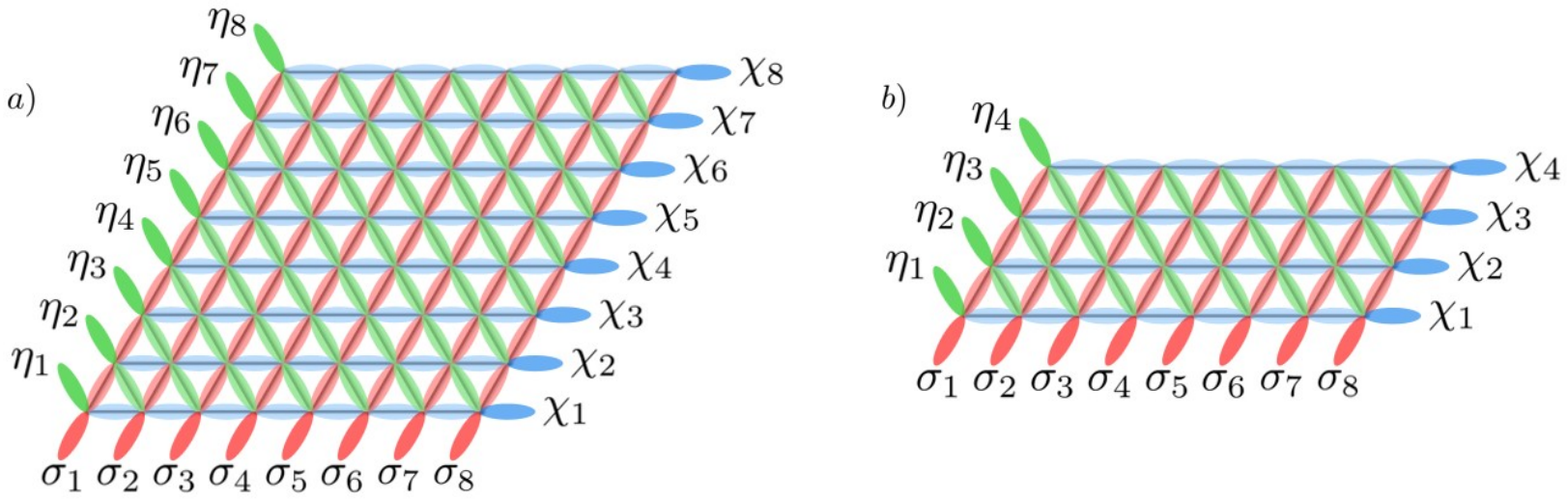}
        \caption{Ising variables along the boundary of (a) an 8$\times8$ cluster and (b) an $8\times4$ cluster. In both cases, the number of distinct $\eta$ variables is $L_c = \mathrm{hcf}(L_a,L_b)$.}
        \label{fig.I_8x8_8x4}
    \end{figure*}

    We derive the energy of a configuration in terms of boundary variables, as given by Eq.~\eqref{eq.E_Ising}. We assign Ising variables $\sigma_i=\pm1$ at the lower boundary of the lattice, for bonds parallel to $\hat{b}$. Similarly, we assign Ising variables $\eta_k=\pm1$ at the left boundary for bonds in the $\hat{b}-\hat{a}$ direction. Ising variables $\chi_j=\pm1$ are assigned at the right boundary for bonds in the $\hat{a}$ direction. When a variable is $+1$, the bond at the boundary has a dimer, and when the variable is $-1$, the bond is empty. Fig.~\ref{fig.I_8x8_8x4} shows examples of lattice configurations with boundary variables. By the second corollary in Sec.~\ref{sec.configspace}, the presence or absence of a dimer at the boundary determines all bonds along the line. It follows that the number of hourglass
    motifs is completely determined by the boundary variables. This is described in Eq.~\eqref{eq.Nhourglass} of the main text. Below, we express the number of $\Psi$ motifs in terms of the boundary variables.

    Each site of the triangular lattice lies at the intersection of three lines that run to the boundary. We associate three integers with each site, $i$, $j$ and $k$, identifying the corresponding line in each direction. The local bond configuration at the site is determined by the Ising variables defined at the end of each line, i.e., by $\sigma_i$, $\chi_j$ and $\eta_k$.
    We define $n_{\Psi}(i,j,k)=0,1$ as the number of $\Psi$ motifs at a given site. By the first corollary in Sec.~\ref{sec.configspace}, we write $n_{\Psi}(i,j,k)=1-n_{Y}(i,j,k)$, where $n_{Y}(i,j,k)$ is the $Y$-motif number. By considering the dimer/blank assignment at bonds attached to a site, it can be shown that
    \begin{equation}
        n_Y(i,j,k)=\begin{cases}
    \frac{1}{4}(\chi_j-\eta_k)(\chi_j-\sigma_i),& \text{if } k \text{ is odd}\\
    \frac{1}{4}(\eta_k-\chi_j)(\eta_k-\sigma_i),              & \text{if } k \text{ is even}
        \end{cases}\;. \label{eq.n_Y}
    \end{equation}
    The total number of $\Psi$ motifs on an $L_a$ by $L_b$ triangular lattice comes out to be 
    \begin{align}
    N_{\Psi}&=
    \sum_{\text{all sites}}n_{\Psi} \nonumber \\
    &=
    \begin{multlined}[t]
    \sum_{\text{all sites}}\left[\left(\frac{1+(-1)^k}{2}\right)n_{\Psi}(i,j,k_{\text{even}})\right.\\
    \left. +\left(\frac{1-(-1)^k}{2}\right)n_{\Psi}(i,j,k_{\text{odd}})\right]. \label{eq.N_P1}
    \end{multlined}
    \end{align}
    Using Eq.~\eqref{eq.n_Y} in Eq.~\eqref{eq.N_P1},
    \begin{equation}
    N_\Psi=\sum_{\text{all sites}}\left[1-\frac{1}{4}\left[1-\chi_j\eta_k+(-1)^k\chi_j\sigma_i-(-1)^k\sigma_i\eta_k\right]\right]. \label{eq.N_P2}
    \end{equation}
    We now note that $i$, $j$ and $k$ are not independent. As the triangular lattice is two-dimensional, we may choose $i$ and $j$ as independent indices. The third index can be expressed as $k=i+j-1$, as can be seen from Fig.~\ref{fig.I_G}. We use this relation in the term proportional to $\chi_j \sigma_i$ to obtain
        \begin{multline}
        N_\Psi=\frac{3}{4}L_aL_b+\frac{1}{4}\left[\sum_{\text{all sites}}\chi_j\eta_k+\sum_{\text{all sites}}(-1)^{i+j}\chi_j\sigma_i\right.\\ +\left.\sum_{\text{all sites}}(-1)^k\sigma_i\eta_k\right]. \label{eq.N_P3}
        \end{multline}
    The sum over all sites can be expressed as a sum over any two indices with appropriate limits. 
    In the term proportional to $\chi_j \sigma_i$, we sum over $i$ and $j$. We must take $1\leq i \leq L_a$ and $1\leq j \leq L_b$ in order to cover all $L_aL_b$ sites. For
    the term proportional to $\sigma_i \eta_k$, we sum over $i$ and $k$. We run $1\leq i \leq L_a$ and $1\leq k \leq L_b$ in order to cover all sites. However, there are only $L_c$ unique values of $k$ (as argued in Appendix~\ref{app.lines}).
    This allows us to sum over $k$ up to $L_c$ and then scale by $L_b/L_c$ to cover all sites. We use a similar approach to sum over $j$ and $k$ in the term proportional to $\chi_j\eta_k$. We set $1\leq j \leq L_b$ and $1\leq k \leq L_c$, scaling by $L_a/L_c$ to account for repetitions. Using these ideas, Eq.~\eqref{eq.N_P3} then becomes
    \begin{multline}
        N_\Psi=\frac{3}{4}L_aL_b+\frac{1}{4}\left[\left(\sum_{j}^{L_b}\chi_j\right)\frac{L_a}{L_c}\sum_{k}^{L_c}\eta_k\right.\\ +\left.\left(\sum_{i}^{L_a}(-1)^{i}\sigma_i\right)\sum_{j}^{L_b}(-1)^{j}\chi_j\right.\\ +\left.\left(\sum_{i}^{L_a}\sigma_i\right)\frac{L_b}{L_c}\sum_{k}^{L_c}(-1)^k\eta_k\right], \label{eq.N_P4}
    \end{multline}
    We use the definitions \eqref{eq.MDefs} and \eqref{eq.ADefs} to rewrite Eq.~\eqref{eq.N_P4} and arrive at Eq.~\eqref{eq.NPsi} of the main text. 
    \newpage
    
\section{Phases in terms of the boundary theory}
    \label{app.IsingPhases}
    In the main text, at the end of Sec.~\ref{sec.effective}, the hourglass phase has been described in terms of the boundary Ising variables. Here, we describe other phases from the phase diagram in the same terms. 
    The effective boundary theory is framed in terms of magnetisation and staggered-magnetisation values.
    For later convenience, we note that magnetisation of an Ising chain is maximum for the configuration $\{+1,+1,+1,+1,\dots\}$ and minimum for $\{-1,-1,-1,-1,\dots\}$. In both cases, staggered-magnetisation is necessarily zero. Similarly, staggered-magnetisation is maximum for $\{-1,+1,-1,+1,\dots\}$ and minimum for $\{+1,-1,+1,-1,\dots\}$. In both cases, magnetisation is necessarily zero.
    \begin{enumerate}[label=(\roman*),wide, labelwidth=!, labelindent=0pt]
        \item \textbf{Rhombus stripes:} In this phase, the system maximizes the number of $\Psi$'s and minimizes that of hourglasses. In order to achieve this, it treats one of the three Ising chains, say \{$\chi$\}'s, differently. It chooses one of four possibilities for the chosen chain: $M_\chi=\pm L_b$ or $A_\chi=\pm L_b$. For the other two chains, there is a single two-fold choice. We must have $M_\sigma=\pm L_a$ and $A_\eta=\pm L_c$, with the \textit{same} sign for $M_\sigma$ and $A_\eta$. With these Ising configurations, Eq.~\eqref{eq.NPsi} yields $N_{\Psi}=L_aL_b$, maximizing the $\Psi$-motifs, while Eq.~\eqref{eq.Nhourglass} yields $N_{\hourglass}=0$. The 24 symmetry-related rhombus stripe configurations are generated by a three-fold choice in picking one of the Ising chains, a four-fold choice for the chosen chain, followed by a two-fold choice for the other two chains combined. The initial three-fold choice amounts to maximizing one of three contributions in Eq.~\eqref{eq.NPsi}, as all three cannot be simultaenously maximized.

        \item \textbf{Stair network phase:} 
        The system minimizes both $\Psi$'s as well hourglasses here. To do this, it treats one of the Ising chains differently. As with the rhombus stripe, there are four possibilities for the chosen chain. For example, if \{$\chi$\}'s are picked, the system chooses $M_\chi=\pm L_b$ or $A_\chi=\pm L_b$. With the other two chains, a single two-fold choice is made. In the example with \{$\chi$\}'s picked, we must have $M_{\sigma}=\pm L_a$ and $A_{\eta}=\mp L_c$, with $M_{\sigma}$ and $A_{\eta}$ having \textit{opposite} signs. With these assignments, Eq.~\eqref{eq.NPsi} gives $N_{\Psi}=L_aL_b/2$ and Eq.~\eqref{eq.Nhourglass} gives $N_{\hourglass}=0$.

        \item \textbf{Staircase/chevron phase:} The boundary variables are chosen in close analogy with the stair network phase. The only difference lies in assigning values to the chosen Ising chain. Here, we assign values such that magnetisation and staggered magnetisation vanish. 
        If \{$\chi$\}'s are chosen, we set $\{\chi_j\}=\{+1,+1,-1,-1,\dots\}$ so that  $M_\chi=0$ and $A_\chi=0$. Note that this also corresponds to a four-fold choice as we may translate $\chi$ values without altering $M_\chi$ and $A_\chi$. 
        These boundary configurations yield $N_\Psi=L_aL_b/2$ and $N_{\hourglass} =L_aL_b/2$.

        \item \textbf{Staggered rhombus stripes:} As with the staircase/chevron phase, we pick one set of Ising variables, say \{$\chi_j$\}, and assign values such that $M_\chi=0$ and $A_\chi=0$. The other two sets of variables \{$\sigma_i$\} and \{$\eta_k$\} are chosen to ensure $M_{\sigma}=\pm L_a$ and $A_{\eta}=\pm L_c$, with $M_{\sigma}$ and $A_{\eta}$ having the \textit{same} sign. This yields $N_\Psi=L_aL_b$ and $N_{\hourglass}=L_aL_b/2$.
        
    \end{enumerate}
    
\end{document}